\documentstyle[12pt]{article}

\newcommand{\dis}{\displaystyle}

\newcommand{\AmS}{{\protect\the\textfont2
  A\kern-.1667em\lower.5ex\hbox{M}\kern-.125emS}}

\begin{document}
\parindent 1.3cm
\thispagestyle{empty}   
\vspace*{-3cm}
\noindent

\def\arccot{\mathop{\rm arccot}\nolimits}
\def\sd{\strut\displaystyle}

\begin{obeylines}
\begin{flushright}
UG-FT-52/94
March 1995
hep-ph/9503466
\end{flushright}
\end{obeylines}

\vspace*{2cm}

\begin{center}
\begin{bf}
\noindent
TAU-PAIR PRODUCTION VIA PHOTON-PHOTON COLLISIONS AT LEP
\end{bf}
  \vspace{1.5cm}\\
FERNANDO CORNET and JOSE I. ILLANA
\vspace{0.1cm}\\
Departamento de F\'\i sica Te\'orica y del Cosmos,\\
Universidad de Granada, 18071 Granada, Spain\\
   \vspace{2.5cm}

{\bf ABSTRACT}

\parbox[t]{12cm}{\indent
We point out that the cross-section for the process
$e^+ e^- \to e^+ e^- \tau^+ \tau^-$ at LEP is large enough to allow
for a study of the anomalous electromagnetic couplings of the $\tau$ lepton.
We show that the present bounds on the magnetic dipole moment can be
improved and that competitive bounds can be obtained for the electric
dipole moment using the data taken from 1992 to 1994.
Finally, we briefly discuss the improvements that can be obtained at LEP II.}
\end{center}

\newpage

The results obtained by the four LEP experiments since they started taking
data have allowed for precision studies
of $Z$-boson physics that have confirmed the Standard Model predictions
at the level of a few per mil \cite{SCHAILE}. This was, certainly, the
main purpose of LEP and, up to now, all the efforts have been focussed in
this direction. However, LEP has other physics potentials that should be
exploited. In particular, two-photon processes have large cross-sections
that might be only one order of magnitude smaller than the
corresponding cross-sections at the $Z$ peak.
In this letter we want to discuss the possibility of studying the
electromagnetic anomalous couplings of the $\tau$-lepton via the process
\begin{equation}
  \label{PROCESS1}
e^+ e^- \to e^+ e^- \tau^+ \tau^-
\end{equation}
at LEP with the data already in tape. We will also give some estimations
of the sensitivity of LEP II to these anomalous couplings.
A similar analysis for
heavy ion colliders has been presented in \cite{ACI}.

The $\tau$-lepton couplings can offer an interesting window to physics
beyond the Standard Model. Although the bounds one can expect to obtain
for these couplings will never be as strong as the ones already available
for the other leptons \cite{EMU}, they may be theoretically more relevant.
Very often the new theories or models predict values for the anomalous
couplings that are enhanced for higher values of the particle mass, making
the $\tau$ the ideal candidate among the leptons to observe these new
couplings.

Two-photon processes offer several advantages to study the electromagnetic
tau couplings at LEP. First of all, this process is free from the
uncertainties originated by possible, anomalous $Z \tau \overline{\tau}$
couplings \cite{BERNABEU}. Moreover, since photons are almost real and
the invariant mass of the $\tau$-pair is very small, we expect the effects
of unknown form-factors to be negligible. Finally, the relevant kinematical
region in the two-photon process is complementary to
the one at PETRA and LEP (in processes mediated by the $Z$), where the
present bounds have been obtained.

The magnetic and electric dipole moments of the tau can be easily
introduced adding two terms to the Standard Model $\tau \overline{\tau} \gamma$
vertex \cite{IZ}:
\begin{equation}
  \label{VERTEX}
-ie\overline{u}(p^\prime)
\left( F_1 (q^2) \gamma^\mu
      + i F_2(q^2) \sigma^{\mu \nu} \dis{q_\nu \over 2 m_\tau}
+ F_3(q^2) \gamma_5 \sigma^{\mu \nu} \dis{q_\nu \over 2 m_\tau}
                    \right) u(p) \epsilon_\mu (q),
\end{equation}
where $ \epsilon_\mu (q) $ is the polarization vector of the photon with
momentum $q$, $F_1(q^2)$ is related to the electric charge, $ e_\tau = e
F_1(0)$,
and $F_{2,3}$ are the form factors related to
the magnetic and electric dipole moments, respectively, through
\begin{equation}
  \label{MOMENTS}
 \mu_\tau = {e (F_1(0)+F_2(0)) \over 2 m_\tau}  \quad ; \quad
  d_\tau = -{e F_3(0) \over 2 m_\tau}.
\end{equation}
In the Standard Model at tree level, $F_1(q^2) =1$ and $F_2(q^2) = F_3(q^2)
=0$.
It should be noted that the $F_2$ term behaves under $C$ and $P$ like the
Standard Model one, while the $F_3$ term violates $CP$. We will restrict our
discussion to tree level. Standard Model radiative corrections can be
calculated and the data can be corrected for their effects.

Bounds on the values of $F_2$ and $d_\tau$ have been obtained
from the study of the $\tau$ angular distribution in
$e^+ e^- \to \tau^+ \tau^-$ at PETRA:
\begin{equation}
   \label{BOUNDS}
\begin{array}{lc}
|F_2| \leq 0.023 & {\hbox{\cite{SILVERMAN}}} \vspace{0.2cm} \\
|d_\tau| \leq 1.6 \times 10^{-16} \; e \, cm  & {\hbox{\cite{PACO}}} .
\end{array}
\end{equation}
These bounds, however, neglect the effects of the form factors from
$q^2=0$ to $\sim 1.5 \times 10^3 \; GeV^2$, where the measurements were taken.
Grifols and M\'endez proposed to look for deviations from
the Standard Model predictions  in the cross-section for
$e^+ e^- \to \tau^+ \tau^- \gamma$ \cite{GRIFOLS}.
Using this method and assuming a
standard $Z \tau \overline{\tau}$ coupling the bounds \cite{LOHMAN}
\begin{equation}
   \label{BOUNDLEP}
      \begin{array}{c}
|F_2(0)|  \leq 0.23 \vspace{0.2cm}\\
|d_\tau|  \leq 1.3 \times 10^{-15} \; e \, cm
      \end{array}
\end{equation}
have been obtained. Recently, much more stringent bounds have been obtained
by Escribano and Mass\'o \cite{ESCRIBANO}
\begin{equation}
   \label{BOUNDLEP2}
|F_2| \leq 0.009  \qquad ; \qquad |d_\tau| \leq 5 \times 10^{-17} \; e \, cm.
\end{equation}
These bounds arise from the good agreement between the Standard Model
predictions and the experimental data obtained at LEP for the decay
width $\Gamma (Z \to \tau^+ \tau^-)$. What it is actually measured, thus,
is the coupling $Z\tau \overline{\tau}$ that can be related to the one
we are concerned with, $\gamma \tau \overline{\tau}$, requiring an
$SU(2) \times U(1)$ gauge invariance linearly realized. Therefore,
the bounds in Eq. (\ref{BOUNDLEP2})
have more theoretical input than the previous ones and the ones that
can be obtained in photon-photon collisions.

The value of the total cross-section in the Standard Model, using the
equivalent photon approximation \cite{BUDNEV}, is
\begin{equation}
\label{CROSS-SECTION}
\sigma (e^+ e^- \to e^+ e^- \tau^+ \tau^-) = 0.303 \; nb .
\end{equation}
Since the collected integrated luminosity from 1992 to 1994 is
$L = 100 \; pb^{-1}$ per experiment, the total number of $\tau$-pairs
produced during these years in no-tag two-photon collisions is $31000$.
Due to the effective $\gamma \gamma$ luminosity, the invariant mass of
these $\tau$-pairs is strongly peaked near threshold. This has two important
consequences. First of all, it allows a clear cut
experimental separation from the $\tau \overline{\tau}$ events produced
in $Z$ decays. Indeed, a cut such as
$M_{\tau \overline{\tau}} \leq 20 \; GeV$, would only
discard a $1.5 \%$ of the two-photon events. Second, it is well known that the
introduction of the anomalous couplings leads to unitarity violations
that do not appear in the Standard Model due to gauge cancellations. Thus,
any realistic theory that produces non-zero values for the couplings
$F_2$ and $F_3$, will also produce some modulating form-factors that will
dampen the contributions from these couplings at large energies. The effects
of these form-factors, however, are expected to be small near threshold,
so we will neglect them.

The dependence of the total cross-section with $F_2$ at LEP, assuming
$F_3 = 0$, is shown in Fig. 1. The cross-section
is a quartic polynomial in $F_2$, but the smallness of
the relevant values of $F_2$ allows for a very good linear approximation as
can be shown in the figure. The dotted line corresponds to the central
value of the Standard Model cross-section (Eq. \ref{CROSS-SECTION})
and the dashed line shows the statistical error per experiment assuming
a $\tau$-pair detection efficiency $\epsilon = 10 \%$. With these assumptions
the bound
\begin{equation}
 \label{F2BOUND}
|F_2(0)| \leq 6.5 \times 10^{-3}
\end{equation}
can be obtained. This bound is stronger than all the previous ones and it is
free of the theoretical assumption that led to the result in Eq.
(\ref{BOUNDLEP2}).
We can exploit the approximate linear dependence on $F_2$
of the cross-section to obtain a simple expression for the bound in
$F_2$ as a function of the total integrated luminosity and the efficiency in
the $\tau$-pair detection:
\begin{equation}
  \label{EQF2}
|F_2(0)| \leq {0.02 \; pb^{-1/2} \over \sqrt{L\epsilon}},
\end{equation}
where $L$ is expressed in $pb^{-1}$ and the efficiency $\epsilon$ is in
per cent.

The behavior of the cross-section as a function of $d_\tau$ is shown in Fig. 2
with the same parameters and notation as in the previous figure. From the
figure
we can read the bound for the electric dipole moment of the $\tau$ that
can be reached at LEP
\begin{equation}
 \label{DBOUND}
|d_\tau| \leq 2.6 \times 10^{-16} \; e \, cm .
\end{equation}
This is a factor of two worse than the bound from PETRA, although it is
measured in a different kinematical region and it is, thus, complementary to
(\ref{BOUNDS}).
The main difference with the magnetic moment case is that, due to
the $CP$-violating nature of the electric dipole moment term, the interference
with the Standard Model cancels. The dependence of the cross-section with
$F_3$ is, thus, quadratic in a very good approximation, hence the
comparatively lower sensitivity of our process to the electric dipole moment
than to the magnetic one. In any case, we can also obtain a simple
expression for the bound in terms of the integrated luminosity and the
efficiency as Eq. (\ref{EQF2}):
\begin{equation}
  \label{EQF3}
d_{\tau}^2 \leq {2.3 \times 10^{-31} \; pb^{-1/2} \over \sqrt{L\epsilon}}
      \; (e \, cm)^2.
\end{equation}

The previous bounds cannot be improved studying the $p_T$ distribution of
the $\tau$-pairs. The only difference in the symmetry properties between the
Standard Model and the magnetic dipole moment term is the helicity-change
introduced by the latter. However, since the $\tau$'s are produced near
threshold, where the mass is of the same order as the photon-photon
center of mass energy, this difference does not generate a measurable
effect in the $p_T$ distribution. With respect to the electric dipole
moment term, the cancellation of the interference with the Standard Model
hides its $CP$-violating nature, so the differences introduced in the
$p_T$ distribution are again negligible.

A different way to look for effects of the electric dipole moment is to study
$CP$-odd observables. Since the initial two-photon state, when adding over
all possible initial configurations, transforms into itself under $CP$, such
observables are a clear signature for $CP$-violation and are expected
to cancel at tree level in the Standard Model. In the photon-photon
center of mass system the $\tau$ and $\overline{\tau}$ momenta remain
unchanged under $CP$ transformations, while their spins are interchanged.
So, in order to isolate a $CP$ violating term, we have to study the production
of polarized taus.\footnote{The expression for polarized $\tau$-pair production
cross-section can be found in \cite{ACI2}.}
Another, more technical, argument in this direction is
that the $CP$ violating term contains the Levy-Civita tensor
$\epsilon_{\mu \nu \alpha \beta}$, so we need four independent vectors to
contract with this tensor, requiring the introduction of the final state
polarization vectors. A complete list of the observables
involving the momenta and polarization of the $\tau$ and $\overline{\tau}$
was given in \cite{BERNREUTHER1}. These observables require the measurement
of the $\tau$ and $\overline{\tau}$ momenta and their polarization analysis
via their decays. However, in two-photon collisions, due to the lack of
knowledge of the center of mass energy and the rapidity in each event and
the presence of the neutrinos in the $\tau$ decay products, the momenta of the
$\tau$ and $\overline{\tau}$ cannot be reconstructed.

We have studied exclusive, one-prong decay modes of the $\tau$, i.e.
\begin{equation}
  \label{PROCESS}
e^+ e^- \to e^+ e^- \tau^+ \tau^- \to e^+ e^- a^+ b^- \nu_\tau
                                      {\overline{\nu}}_\tau X ,
\end{equation}
where $a^+$, $b^-$ are the charged particles in the $\tau^+$, $\tau^-$
decay products, respectively, and X might be any other neutral particle.
The observables will be constructed only in terms of the momenta of the
final, observed charged particles.  The polarization-analyzing power of the
decay channels is characterized by a constant
$\alpha_a$ (with $| \alpha_a | \leq 1$)
which is shown in Table 1 for the most relevant decay channels, together
with their respective branching ratios \cite{BERNREUTHER2}.
The most sensitive observable turns out to be:
\begin{equation}
 \label{OBSERVABLE}
\hat{T}_{33} = (\hat{q}^+ - \hat{q}^-)_3
        {(\hat{q}^+ \times \hat{q}^-)_3 \over |\hat{q}^+ \times \hat{q}^-|},
\end{equation}
where $\hat{q}^\pm$ is the unit vector in the direction of the corresponding
final state charged particle. Since the $\tau$ couplings to the photon are
vector-like, the expectation values of this observable for the channel in
Eq. (\ref{PROCESS}) is related to the pion-pion channel through:
\begin{equation}
 \label{RELATION}
<\hat{T}_{33}>_{a^+ b^-} = \alpha_a \alpha_b <\hat{T}_{33}>_{\pi^+ \pi^-}.
\end{equation}
{}From the values at Table 1 it is easy to see that the best channel to
look for $CP$-violating effects in this observable is:
$ e^+ e^- \to e^+ e^- \tau^+ \tau^- \to e^+ e^- \pi^+ \pi^- \nu_\tau
{\overline{\nu}}_\tau$. Using this channel the bound that can be obtained
is:
\begin{equation}
 \label{BOUNDEDMCP}
|d_\tau| \leq 2.3 \times 10^{-15} \; e \, cm .
\end{equation}
This is one order of magnitude higher than the bound that can be obtained from
the total cross-section because of the poor statistics due to the small
$ \tau \to \pi \nu$ branching ratio.

At LEP II the bounds obtained at LEP can be improved because of two reasons.
First of all, contrary to the behavior of the annihilation process, the
cross-section for (\ref{PROCESS1}) grows with the $e^+ e^-$
center of mass energy. Second, the design luminosity of LEP II is larger
than the one of LEP. In summary, the number of $\tau$-pairs that can be
produced in one year at each LEP II experiment with a center of mass energy
$\sqrt{s} = 200 \; GeV$ and an integrated luminosity $ L = 300 \; pb^{-1}$
is $141000$. Thus reducing the statistical error a factor $\sim 2.2$ with
respect to the LEP error. The bounds for the magnetic and electric dipole
moments that can be obtained from the total cross-section, assuming again
an efficiency $\epsilon = 10 \%$ are:
\begin{equation}
  \label{BOUNDSLEPII}
    \begin{array}{c}
|F_2(0)| \leq 2.5 \times 10^{-3} \vspace{0.2cm}\\
|d_\tau| \leq 1.6 \times 10^{-16} \; e \, cm .
\end{array}
\end{equation}
The situation for the $CP$-odd observables is similar to the one we found
at LEP. The bound on $d_\tau$ that can be obtained is sensibly worse than
the one in Eq. (\ref{BOUNDSLEPII}), $d_\tau \leq 10^{-15} \; e \, cm$.

In summary, we have discussed the possibility of using the data that is
already on tape at LEP to look for effects of anomalous magnetic and
electric dipole moments in two photon processes. The cross-section for
$\tau$-pair production in photon-photon collisions at LEP is large
enough to allow for an improvement of the present bounds on the magnetic
dipole moment. The bounds that can be obtained for the electric dipole
moment are a factor $\sim 1.6$ worse than the present bounds. However,
we should stress that these bounds are obtained with real photons and very
near threshold. So, the kinematical region and the obtained bounds
are complementary to the present ones obtained at PETRA in $e^+e^-$
annihilation. Due to the impossibility of reconstructing the $\tau$'s in
this process, $CP$-odd observables have to be constructed using only the
momenta of the $\tau$ decay products. The statistics is, thus, much poorer
than for the total cross-section measurement and the bound that can be
obtained using these observables is worse than the one obtained via the
total cross-section.

We thank F. del Aguila and C. Ma\~n\'a for very helpful discussions and
comments. This work was supported by CICYT under contract AEN94-0936,
by the European Union under contract CHRX-CT92-0004 and by Junta de
Andaluc{\'\i}a.

\newpage

\begin{table}
\begin{center}
\begin{tabular}{ccc}
\qquad Decay Mode \qquad & \qquad $\alpha_a$ \qquad & \qquad BR ($\%$) \qquad
\\
\hline
$\pi^- \nu_\tau$                &       $1$        &    $ 11 $        \\
$\rho^- \nu_\tau$               &       $0.46$     &    $ 23 $        \\
$l^- \nu_\tau \overline{\nu}_l$ &  $-{1 \over 3}$  &    $ 36 $        \\
\end{tabular}
\caption{Values of polarization analyzing constant $\alpha_a$ and branching
ratios for the $\tau$ decay modes considered.}
\end{center}
\end{table}
\vspace*{3cm}

\noindent
{\bf Figure Captions}

\noindent
\begin{description}
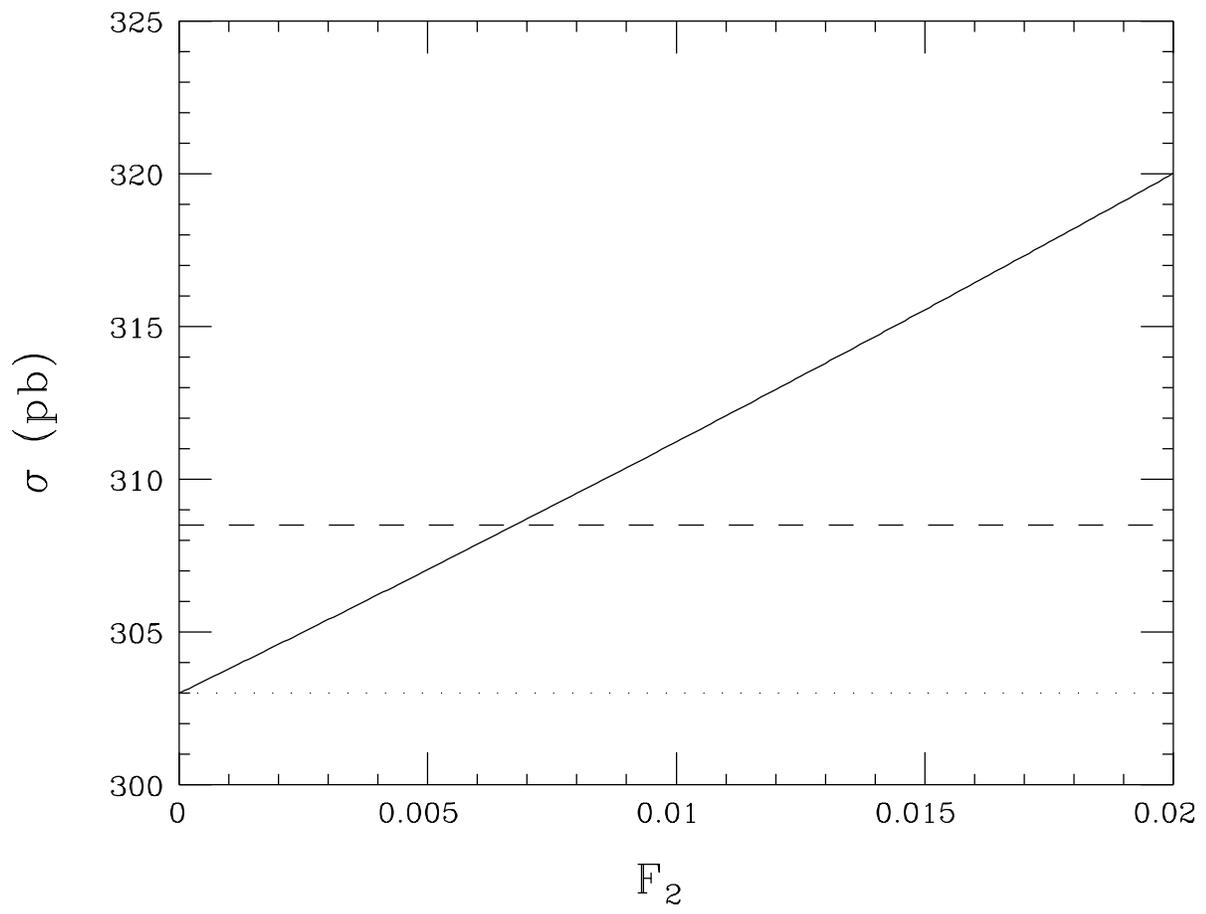
\item[Fig. 1:] Total cross section for the process $e^+ e^- \to
e^+ e^- \tau^+ \tau^-$ as a function of the anomalous magnetic moment, $F_2$
(solid line). The dotted line stands for the Standard model prediction and
the dashed line shows the statistical error assuming an integrated luminosity
of $100 \; pb^{-1}$ and a $10 \%$ efficiency in the detection of the produced
$\tau$-pairs.

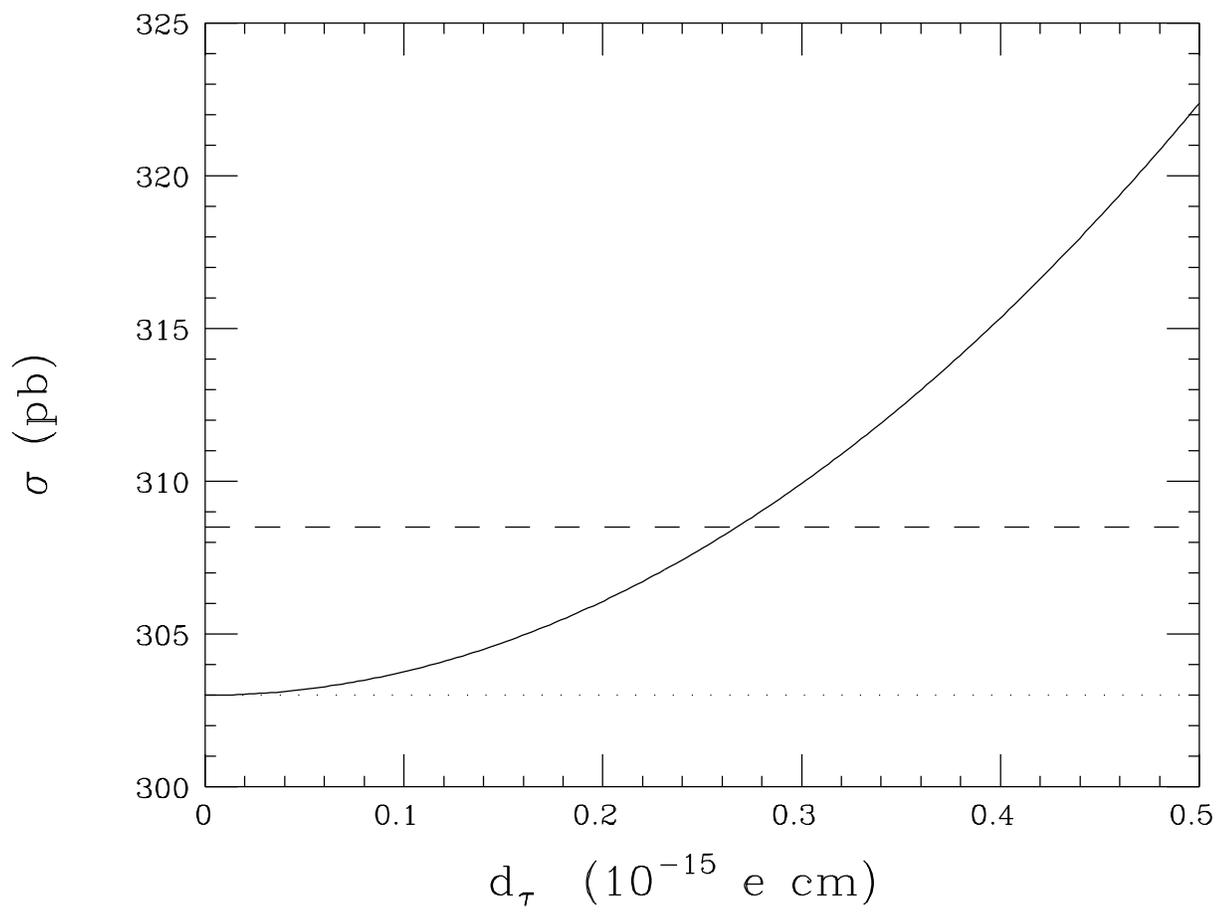
\item[Fig. 2:] Total cross section for the process $e^+ e^- \to
e^+ e^- \tau^+ \tau^-$ as a function of $d_\tau$ with the same conventions as
in the previous figure.
\end{description}

\end{document}